\documentclass{elsart}
\usepackage{graphicx}
\usepackage{amsfonts}
\usepackage{amsmath}
\usepackage{bm}
\usepackage{amssymb}
\usepackage{epsfig}
\journal{Communications in Nonlinear Science and Numerical Simulations}

\theoremstyle{definition}

\theoremstyle{remark}

\numberwithin{equation}{section}

\newcommand{\abs}[1]{\lvert#1\rvert}

\newcommand{\Dql}{D_{\mathrm{QL}}}

\newcommand{\eg}{{\textrm{e.g.\ }}}
\newcommand{\ie}{{\textrm{i.e.\ }}}
\newcommand{\rmd}{ \, {\mathrm{d}}}
\newcommand{\rme}{{\mathrm{e}}}
\newcommand{\rmi}{{\mathrm{i}}}

\newcommand{\EE}{{\mathbb E \, }}

\newcommand{\PP}{{\mathbb P}}

\newcommand{\ZZ}{{\mathbb Z}}

\begin{document}
\begin{frontmatter}

\title
      {Nonquasilinear evolution of particle velocity in
      incoherent waves with random amplitudes
       }

\author{Yves Elskens}
\address{Physique des interactions ioniques et mol\'eculaires,
         UMR 6633 CNRS--universit\'e de Provence,
         Aix-Marseille Universit\'es, \\
         Equipe turbulence plasma, case 321 campus Saint-J{\'e}r{\^o}me,
         FR-13397 Marseille cedex 13 \\
         {\tt yves.elskens@univ-provence.fr}}
\thanks{Supported in part by a CNRS temporary position during this work.}

\date{\today}

\begin{abstract}
  The one-dimensional motion of $N$ particles in the field of many incoherent waves
  is revisited numerically.
  When the wave complex amplitudes are independent,
  with a gaussian distribution,
  the quasilinear approximation is found to always overestimate transport
  and to become accurate in the limit of infinite resonance overlap.
\end{abstract}
\end{frontmatter}

The motion of a particle in the field of many waves is a fundamental process in
classical physics \cite{DG03,DoveilMacor06,Tsunoda91}, usually discussed in the first
chapters of plasma physics textbooks. For modeling purposes, this process can be
described by the hamiltonian model
\begin{equation}
  H = \frac{v^2}{2} + \sum_{m=1}^M A_m \cos (k_m x - \omega_m t - \varphi_m)
  \label{Hns}
\end{equation}
where the particle has position $x$ and momentum $v$ (normalizing mass to 1 and writing
$A_m = q E_m/k_m$). The wave field has $M \gg 1$ waves, with a smooth dispersion
relation associating a wavenumber $k_m$, a pulsation $\omega_m$ and a phase velocity
$v_m = \omega_m / k_m$ to each wave -- usually determined by fixed properties of the
environment, such as the geometry of the domain where waves propagate (then the $k_m$
are discrete). The actual spectrum of the wavefield, given by the complex amplitudes
$A_m \rme^{\rmi \varphi_m}$, is more easily tuned by the experimenter or affected by
simple changes in the environment, \eg excited antennas.

The dynamical system approach to this problem would be to prescribe a single choice for
each wave complex amplitude. However it would be quite exceptional to control all waves
(though this is \eg the assumption underlying the standard map with $A_m = A_0$,
$\varphi_m = \varphi_0$ for all waves -- see \cite{BE98Std} for a discussion).

Because the waves have different frequencies and velocities, it is generally unrealistic
to assume their phases to be correlated. Their intensities are more easily observed, but
both in nature and in the laboratory the accumulation of statistics on waves often
involves only their average power spectra, not the detailed intensity data for each
measurement run. Besides, if the waves are excited by a noisy source, as occurs in the
self-consistent dynamics of particles and waves \cite{DoxasCary,EEbook}, one cannot
expect a systematic reproduction of individual wave data among several experiments (this
is the very meaning of a ``noisy source'').

We thus assume here that these complex amplitudes are random data, and investigate the
statistics of the particle motion in the resulting time-dependent random field. This
dynamics is a ``stochastic acceleration problem'' for a ``passive particle'' in weak
plasma turbulence \cite{DG82,Sturrock66,VdEijnden97}, and its understanding is a
prerequisite to a proper analysis of the case where the particle motion feeds back on
the wave evolution \cite{EEbook}.

This hamiltonian generates equations of motion
\begin{eqnarray}
  \dot x & = & v
  \label{dotx} \, ,
  \\
  \dot v & = & \sum_m k_m A_m \sin (k_m x - \omega_m t - \varphi_m)
  \label{dotv}  \, .
\end{eqnarray}

A simple case for the dispersion relation is
\begin{equation}
  k_m = k_0 , \ \omega_m = 2 \pi (m-M/2) / T  ,
  \label{BEspectrum}
\end{equation}
for some $k_0$, $T$. Then, in the limit $M \to \infty$, the equations of motion yield
for $A_m \rme^{\rmi \varphi_m} = A_0$ (with real $A_0$) the well-known standard map
\cite{BE98Std}. The case where phases $\varphi_m$ are independent random variables
uniformly distributed on the circle $[0, 2 \pi]$ while $A_m = A_0$ is given was
investigated notably by Cary, Escande, Verga and B\'enisti \cite{BE97,CEV,EEbook} and
occurs in the context of the random phase approximation.

An important observation \cite{Chirikov79,Escande85} on the motion of a particle in the
field (\ref{dotv}) is locality in velocity~: the evolution of the particle when it has
velocity $v$ depends only weakly on the waves with a Doppler-shifted frequency $\omega_m
- k_m v$ much larger than their trapping oscillation frequency $2 k_m \sqrt{A_m}$. In
particular, for a two-wave system the resonance overlap parameter
\begin{equation}
  s_{1,2}
  =
  \frac {2 \sqrt{A_1} + 2 \sqrt{A_2}} {\abs{\omega_2 / k_2 - \omega_1 / k_1}}
\label{s12}
\end{equation}
becomes unity when there exists a velocity $u$ for which $\omega_2/k_2 - u = 2
\sqrt{A_2}$ and $u - \omega_1/k_1 = 2 \sqrt{A_1}$ (with $k_1>0$, $k_2 > 0$, $\omega_2 >
\omega_1$). For many waves with overlap parameters $s \gg 1$, the relevant phase
velocity range for waves influencing the particle is a ``resonance box'', with a width
scaling as $A^{2/3}$ \cite{BE97,BE98}.

To the extent that the phases and amplitudes of the waves are independent random
variables, it is tempting to approximate the acceleration (\ref{dotv}) by a white noise,
with amplitude $\sigma = \sqrt{\EE (k_m^2 A_m^2)} = k_m \sqrt{\EE A_m^2}$, where the
relevant mode $m$ is the one with the current particle velocity ($\EE$ denotes the
mathematical expectation, or ensemble average with respect to wave amplitudes and
phases).\footnote{Phases do not appear in $\sigma$ (nor in $s$) because the relative
phase of two waves $\varphi_m + \omega_m t - \varphi_n - \omega_n t$ varies uniformly
over time (hence $\varphi_m - \varphi_n$ can be absorbed in the choice of the time
origin).}

This is the core of quasilinear theory \cite{Drummond62,Pesme94,Romanov61,Vedenov62}.
Mathematically, one then interprets (\ref{dotx})-(\ref{dotv}) as a stochastic
differential equation, after the necessary reformulation to make sense of divergent
series~; this is correct in the case $m \in \ZZ$ for dispersion relation
(\ref{BEspectrum}) with gaussian independent complex amplitudes such that $\EE A_m^2 =
k_0^{-2} \sigma^2$. The particle velocity then has a brownian evolution, so that for $0
< t \leq t' < T$
\begin{equation}
  \EE (v_t - v_0)(v_{t'} - v_0) = 2 \Dql \min(t,t')
  \label{v2tql}
\end{equation}
with the quasilinear diffusion coefficient $\Dql = \sigma^2 T/4$. In particular,
\begin{equation}
  \EE \Delta v_t^2 = {\frac{\sigma^2 T} 2} \, t
  \label{dv2}
\end{equation}
for $\Delta v_t = v_t - v_0$. However, the particle evolution for $t > T$ may show a
strong correlation to its motion for $0 \leq t \leq T$ because the waves are periodic in
time \cite{Elskens07a,ElskensPardoux08}, and the dispersion relation (\ref{BEspectrum})
may generate a strong spatial correlation -- although it just accounts for the fact that
in the limit of a dense spectrum (with a smooth dispersion relation) the relevant waves
acting on a particle at any time also have almost the same wavelength.\footnote{~There
is a large body of literature on the case of incoherent waves with no dispersion
relation. Then the sum $\sum_m$ becomes a double sum $\sum_{m,n}$ and one varies
wavenumbers $k_n$ independently from pulsations $\omega_m$. This space-time stochastic
environment is more noisy that our model and may also be considered to motivate a
quasilinear approximation.}

In this note we compare numerical properties of the velocity evolution for fixed
amplitude with the case of random wave amplitudes~: let $A_m \rme^{\rmi \varphi_m} = C_m
+ \rmi S_m$. In the latter case, the amplitudes are drawn independently in such a way
that their squares have exponential distributions, with equal expectation $\EE A_m^2 =
k_0^{-2} \sigma^2$. Along with uniform distribution of the phases, this is equivalent to
gaussian distribution of the complex amplitudes. Indeed, the probability density
$f_\varphi$ for $\varphi$ is uniform, and the exponential law for $A^2$ yields for $A$
the density at $a>0$
\begin{equation}
  f_A(a) = \frac{\PP(a \leq A < a + \rmd a)}{\rmd a}
  = \frac{ \PP(a^2 \leq A^2 < a^2 + 2 a \rmd a) }{ \rmd a }
  = \rme^{-a^2/\EE A^2} \frac{2 a}{\EE A^2}
  \label{explaw} \, .
\end{equation}
The probability density for $(C,S)$ is thus (noting that $a^2 = c^2 + s^2$ and $\rmd c
\rmd s = a \rmd a \rmd \varphi$)
\begin{equation}
  f_{C,S} (c,s)
  = \frac{f_{A}(a) f_\varphi(\varphi) \rmd a \rmd \varphi}{\rmd c \rmd s}
  = \frac{2 a \rme^{-a^2/\EE A^2}}{2 \pi \, \EE A^2} \frac{\rmd a \rmd \varphi}{\rmd c \rmd s}
  = \frac{\rme^{-(c^2 + s^2)/\EE A^2}}{\pi \, \EE A^2}
  \label{fcs}
\end{equation}
which is the density for a gaussian vector $(C,S)$ with zero expectation and covariance
matrix $  \Bigl(
     \begin{array}{cc}
        \EE A^2 / 2 & 0  \cr
        0 & \EE A^2 / 2
     \end{array}
     \Bigr)$.

\begin{figure}
\centerline{
  \psfig{figure=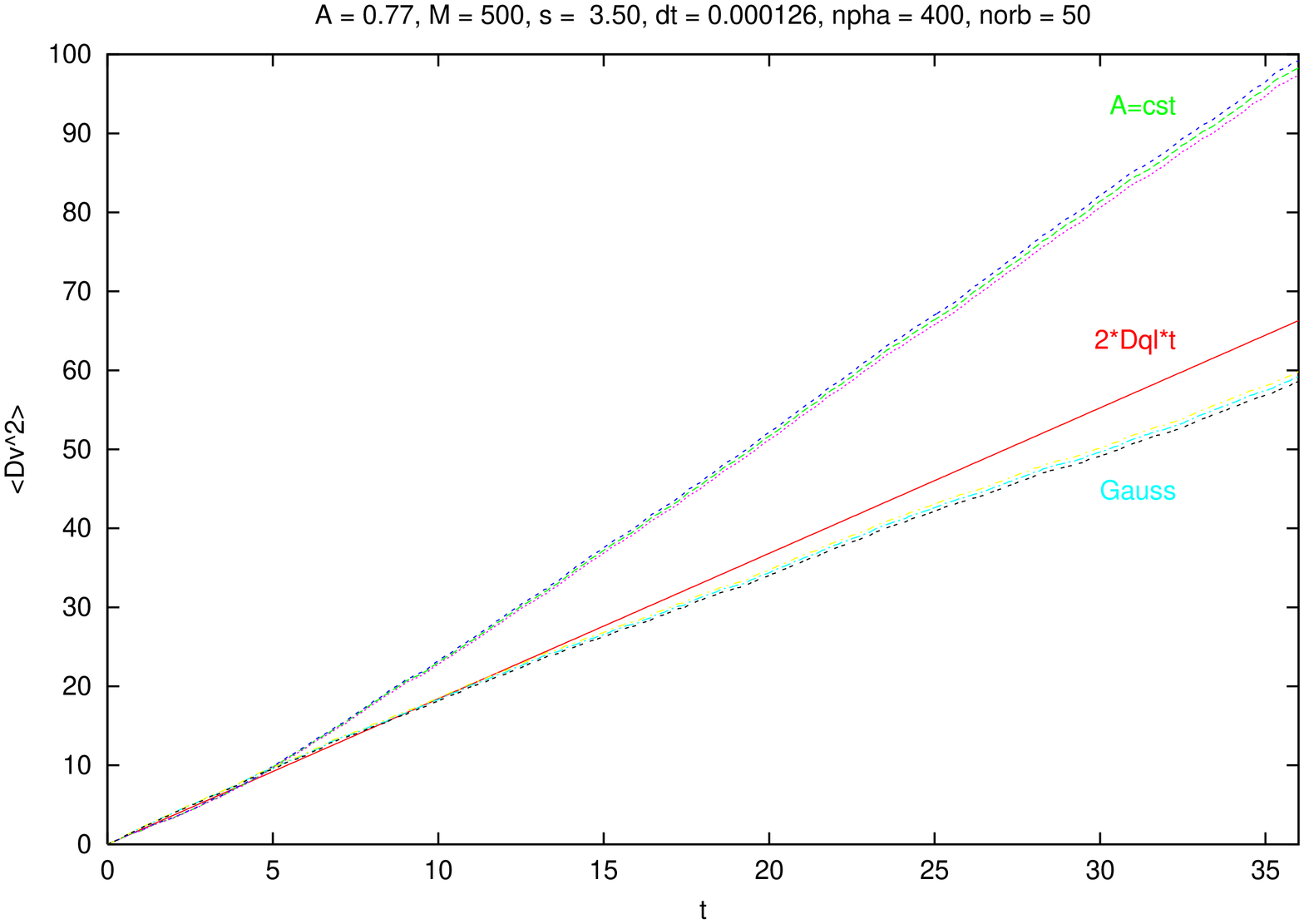,width=8cm,height=13cm,angle=-90}
  }
  \vskip3mm
  \centerline{
  \psfig{figure=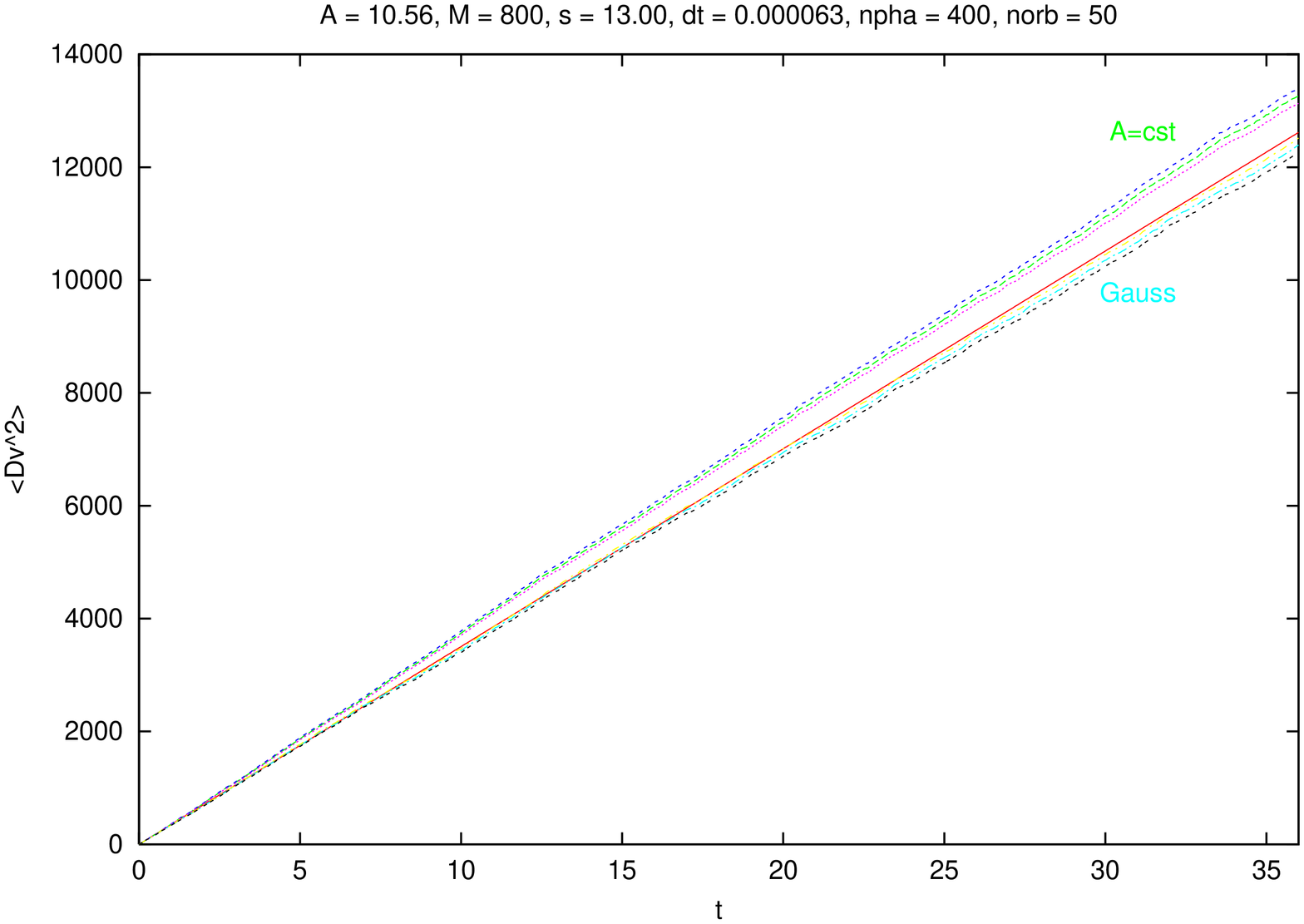,width=8cm,height=13cm,angle=-90}
  }
 \vskip3mm %
\caption{
  Dependence of $\langle \Delta v^2 \rangle$ on time for random amplitudes
  (Gauss) and for equal amplitudes ($A$=cst).
   }
\label{fig1}
\end{figure}

For the simulations, we let $k_0 = 1$ and $T = 2 \pi$ so that $\sigma = s^2/16$. Given
the value of $\sigma$, or equivalently of $s$, we draw at random $N_{\mathrm{pha}} =
400$ sets of wave data. For each set of wave data, we follow the evolution of
$N_{\mathrm{orb}} = 50$ particles, released initially at random points $(x,p)$ in the
strip $0 \leq x \leq 2\pi$, $-0.5 \leq p \leq 0.5$. Trajectories are computed with a
reversible symplectic integrator up to a time, large enough for the motion to exhibit
possible departure from quasilinear statistics, but short enough to ensure that the
particles remain far from the boundaries of the wave spectrum, which have velocity $\pm
M/2$ (where we take $M$ up to 800)~: for $s=13$, the boundaries are more than 3 standard
deviations (which is about $120$ as observed in Fig.~\ref{fig1}) away from the initial
particle velocity. Some trajectories are computed backward from their final point to
control numerical accuracy~: the accumulated error is about $10^{-5}$ for $s=3.5$ for
the runs of Fig.~\ref{fig1} and \ref{fig3}, though it deteriorates rapidly with
increasing $s$. Our calculations for the equal $A$ case reproduce the findings of Refs
\cite{BE97,CEV}.

Statistical averages discussed below (denoted by $\langle \cdot \rangle$) are performed
over all particle and wave data for the same $s$ value. We focus on the second and
fourth moments of particle velocity as functions of time. To the numerical average
$g_2(t) = \langle \Delta v^2 \rangle$ we fit a linear approximation $D_0 + 2
D_{\mathrm{eff}} t$, which defines an effective diffusion coefficient. Such a plot is
displayed in Fig.~\ref{fig1}. The quasilinear prediction $2 \Dql t$ is a straight line.
The numerical average $g_2$ is plotted, along with two lines estimating one standard
deviation on either side, viz.\ $g_2 \pm \langle \Delta v^4 - g_2^2 \rangle^{1/2}$. The
narrowness of the channel so constructed indicates the statistical accuracy of the
plotted function $g_2$.

Specifically, Fig.~\ref{fig1} displays our results for $s=3.5$ and for $s=13$. The
non-overlap regime ($s \lesssim 1$) allows no large scale transport in velocity, because
the particle phase space contains invariant Kolmogorov-Arnol'd-Moser tori. Indeed the
case $s=0$ is integrable (it corresponds to free particle motion). For increasing
overlap parameter value, one expects the transport to become increasingly chaotic, and
indeed the particle dynamics typically has a Liapunov exponent scaling like
 $A^{1/3} \sim s^{2/3}$ for large $s$ \cite{EEbook}. One may then expect
the effective diffusion coefficient to increase towards the quasilinear estimate,
$\Dql$, which corresponds to a pure stochastic particle behaviour. However it has been
found \cite{BE97,CEV,EEbook} that for the equal amplitude, independent phases wave data,
the particle velocity undergoes an enhanced transport, with an effective diffusion
coefficient up to 2.5 times the quasilinear estimate, and the latter becomes accurate
only in the large $s$ limit. It must be noted that this enhanced transport shows up only
after time $T=2\pi$, \ie only on a time scale for which the time periodicity due to the
discrete nature of the wave spectrum significantly affects the particle motion --
actually the motion for short times is essentially quasilinear thanks to the wave
stochasticity \cite{EEbook,EscandeElskens03}. We also checked that the force correlation
function on the particle is essentially zero, except for times near integer multiples of
the period $T$.

\begin{figure}
\centerline{
  \psfig{figure=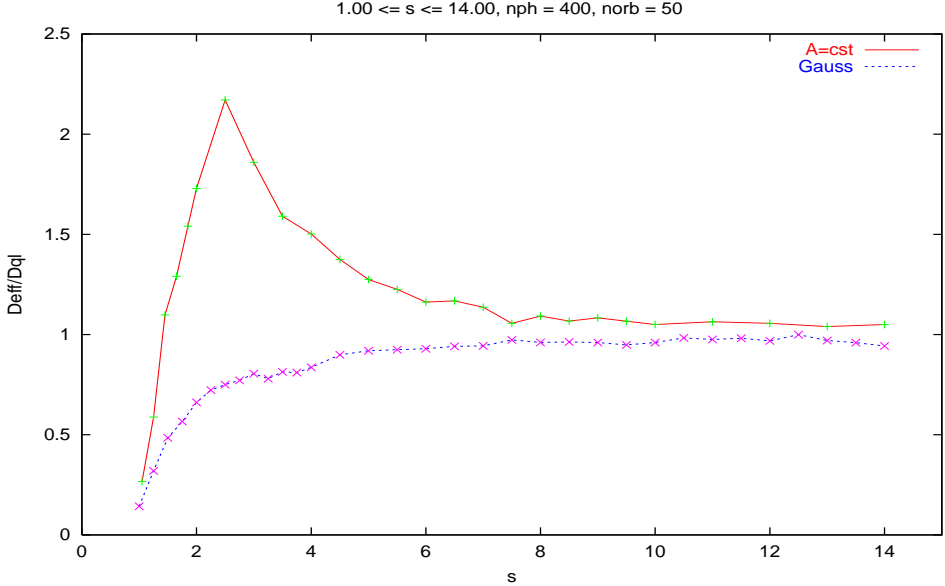,width=8cm,height=13cm,angle=-90}
  }
  \vskip3mm %
\caption{
  Ratio $D_{\mathrm{eff}}/\Dql$ as function of overlap parameter $s$ for random amplitudes
  (Gauss) and for equal amplitudes ($A$=cst). The lines are guides to the eye.
   }
\label{fig2}
\end{figure}

\begin{figure}
\centerline{
  \psfig{figure=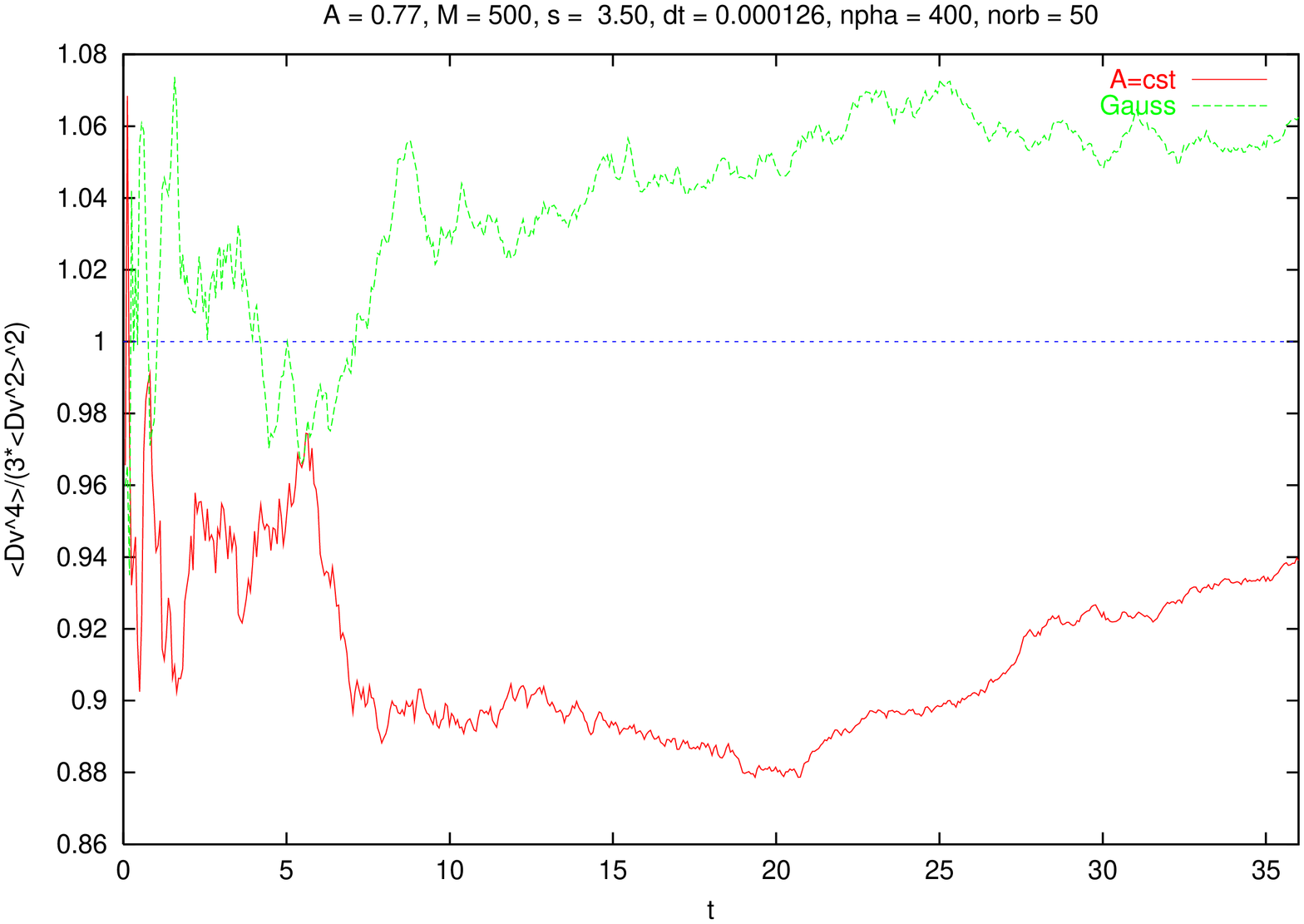,width=8cm,height=13cm,angle=-90}
  }
  \vskip3mm
  \centerline{
  \psfig{figure=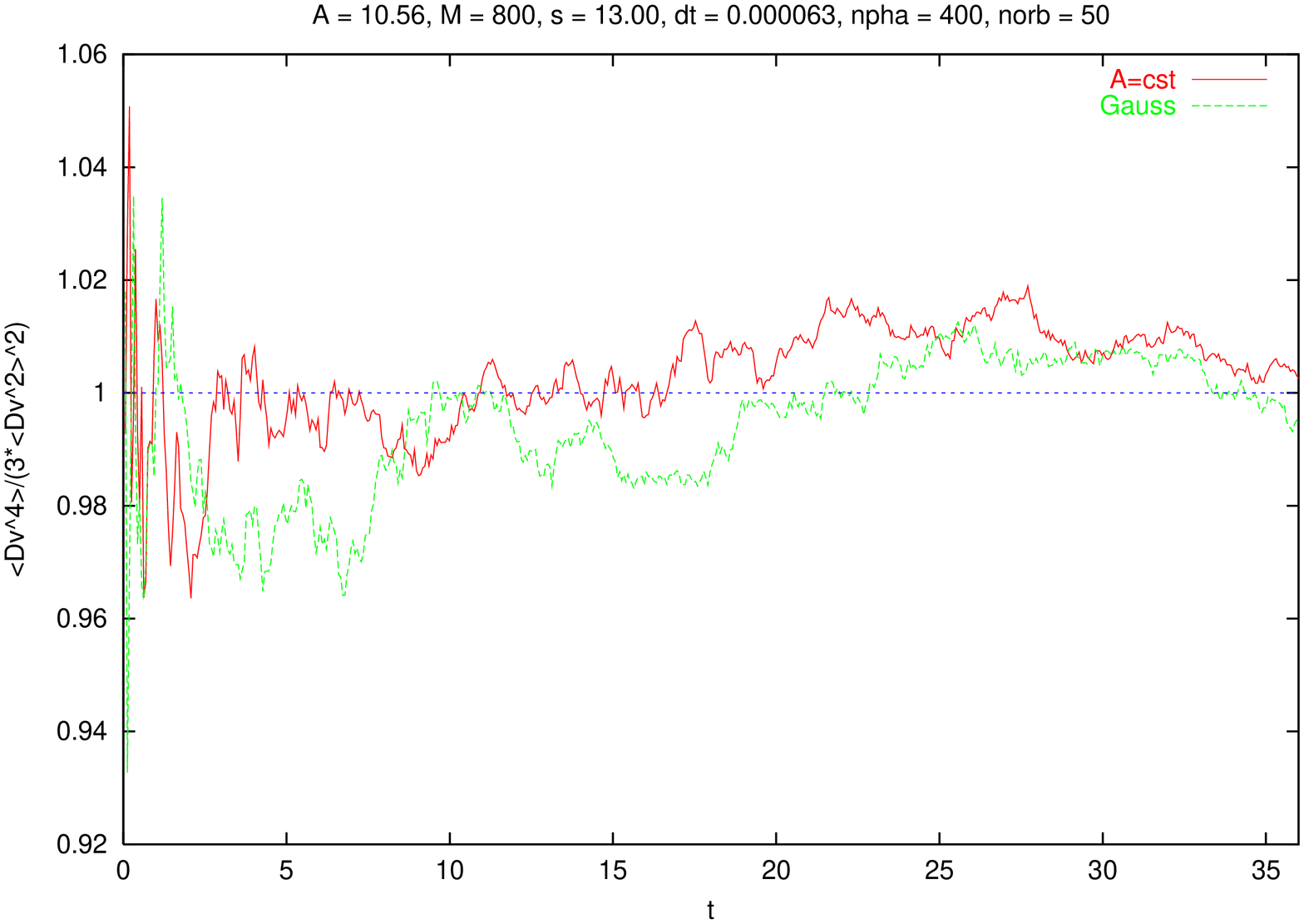,width=8cm,height=13cm,angle=-90}
  }
 \vskip3mm %
\caption{
  Ratio $\langle \Delta v^4 \rangle / (3 \langle \Delta v^2 \rangle^2)$ versus time
  for random amplitudes (Gauss) and for equal amplitudes ($A$=cst).
   }
\label{fig3}
\end{figure}

Our simulations show that the enhanced transport is not observed for independent wave
amplitudes. In the latter case, the velocity variance grows with time at a rate never
exceeding the quasilinear value. As shown in Fig.~\ref{fig2} this observation holds for
the whole range of overlap parameter values sampled, in agreement with the naive
prediction that ``increasing chaos'' should make dynamics more similar to ``pure
noise''.

The noisiness of the particle evolution is also assessed using a higher-moment test. If
the velocity distribution at time $t$ were gaussian, the ratio $\langle \Delta v^4
\rangle / (3 g_2^2)$ would equal 1. Fig.~\ref{fig3} displays numerical evidence that for
the equal wave amplitude case this ratio remains rather below 1 (the velocity
distribution has smaller tails than the gaussian with equal variance), whereas for the
gaussian wave data it is somewhat above 1 (the distribution has stronger tails than the
gaussian with same variance).

For both wave data types the limiting behaviour for $s \to \infty$ is quasilinear, in
agreement with theoretical \cite{BE97,EEbook,EscandeElskens02} and mathematical
\cite{Elskens07a,ElskensPardoux08} arguments.

These results confirm that random phases only are not sufficient to substantiate the
quasilinear approximation for the stochastic acceleration problem. A gaussian wave
spectrum may seem closer to the ideal view of white noise for one period $T$, but over
longer times correlations also build up, driving transport away from the quasilinear
approximation, especially when the overlap parameter has moderate typical values ($1
\lesssim s \lesssim 5$).

Many fruitful discussions with F.~Doveil and D.F.~Escande are acknowledged.


\end{document}